\documentstyle[epsfig,multicol,prl,aps]{revtex}
\ifpreprintsty
 \def\multb{ }
 \def\multe{ }
 \else
 \def\multb{ \begin{multicols}{2}}
 \def\multe{ \end{multicols}}
 \fi

\begin{document}
\draft
\title{Lattice dynamics and reduced thermal conductivity of filled skutterudites }
\author{J. L. Feldman, D. J. Singh, I. I. Mazin}
\address{Code 6391, Naval Research Laboratory, Washington, D.C. 20375-5345}
\author{D. Mandrus and B. C. Sales}
\address{Solid State Division, Oak Ridge
National Laboratory, Oak Ridge, Tennessee 37831}
\date{\today}
\maketitle

\begin{abstract}
The great reduction in thermal conductivity of skutterudites
upon filling the ``void'' sites with Rare Earth ({\it RE}) ions is key to
their favorable thermoelectric properties but remains to be understood.
Using lattice dynamic models based
on first principles calculations, we address the most popular microscopic
mechanism, reduction via rattling ions. The model
withstands inelastic neutron scattering and specific heat
measurements, and refutes hypotheses of an anharmonic {\it RE} potential and
of two distinct localized {\it RE} vibrations of disparate
frequencies. It does indicate a strong
hybridization between bare La vibrations and certain Sb-like phonon
branches, suggesting {\it anharmonic }scattering by {\it harmonic} RE
motions as an important mechanism for suppression of heat conductivity.
\end{abstract}


\multb
The discovery of new high performance thermoelectrics with complex
crystal structures poses an important challenge for theory, which, if
met, could lead to the synthesis of materials with substantially higher
thermoelectric performance. The key is to learn how the contradictory
criteria --- high electrical and low thermal conduction ---
are satisfied. One of the most promising groups of materials
are the filled skutterudites. Skutterudites (MPn$_{3}$:
M=Co,Rh,Ir; Pn=P,As,Sb) can be described as a cubic lattice of M
atoms with 3/4 of the cubes filled by nearly square Pn$_{4}$ rings. They
have reasonable electronic properties for thermoelectric
application, but ordinary values ($\approx $100 mW/cmK at $T$=300K) of
thermal conductivity. It was conjectured by Slack that a great
reduction in thermal conductivity could be achieved by filling the empty
cubes (voids) with rare gas atoms. The idea was that the fillers would
``rattle'' inside the voids, strongly scattering low frequency phonons
but not charge carriers. It was subsequently found that filling
with rare earths (La, Ce) reduces the thermal conductivity by a factor of
more than five at room temperature. \cite{sales,caill}
Essentially, regarding the
thermal conductivity, these crystalline materials may behave like glasses. The
origin of this strange behavior, also seen in some other materials
(clathrates, KBr-KCN), is basically not understood. The occasionally
discussed possibility that such a large reduction of the lattice thermal
conductivity (LTC) is due to scattering by electronic excitations can be
rejected by various arguments based on experimental trends\cite{elph}.
On the other hand, sorting out other mechanisms requires microscopic analysis.
The most notable hypothesis
relates to anomalous dynamics of {\it RE} atoms, namely strongly anharmonic
rattling motion.
It has also been suggested, from experiment, that there
exist two inherent {\it RE} vibrations (or perhaps two different two level
systems (TLS's))
of significantly different frequencies implying
strong anharmonicity of a different nature than simple rattling.
Here, we construct a microscopic model needed to address these issues,
demonstrate that it describes the
experiments, and use it to discuss possible scenarios for the
LTC reduction.

Our lattice dynamical model is based in part
on first principles density
functional calculations, performed by the linearized augmented planewave
(LAPW) method\cite{sing}, as described in Ref.\cite{dsingh}, including
directly calculated forces generated by numerous ($\approx $ 40) sets of
atomic displacements small enough to represent the harmonic coupling
constants along various directions.
Previously, we\cite{FS} used the LDA to obtain
A$_{g}$ and A$_{u}$ vibrational modes for
CoSb$_3$ within a frozen phonon direct method
and combined these 
with results of infrared measurements to develop a valence force
field lattice dynamical model. Here we start with the model
for CoSb$_{3}$, and make
a minimal number of additions to, and adjustments of parameters needed
to reproduce our new LDA results with a reasonable accuracy.
However, we shall first address the shape of the {\it RE} potential well.

\begin{figure}[t]
\centerline{\epsfig{file=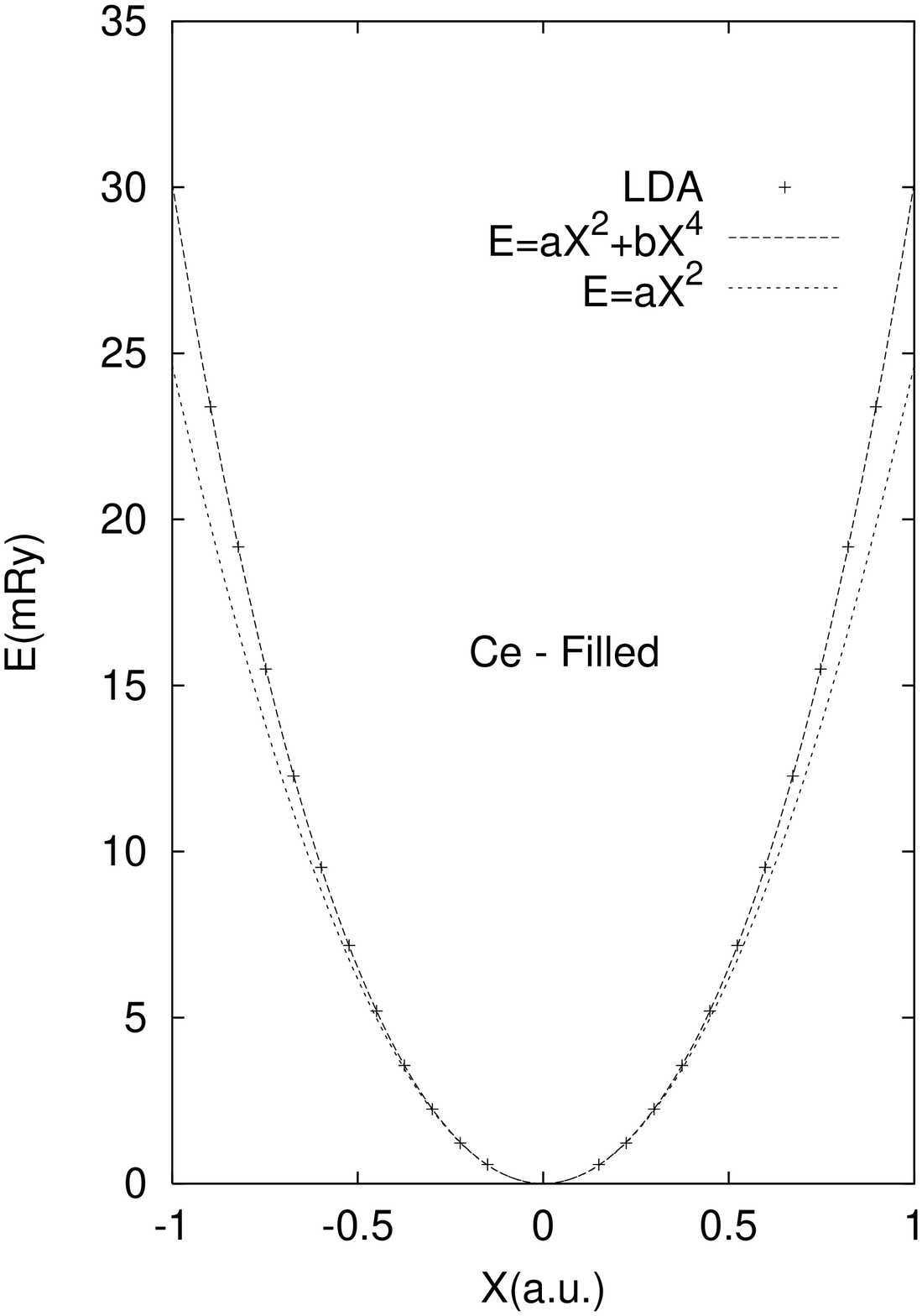,width=0.48\linewidth,height=.48\linewidth}
\epsfig{file=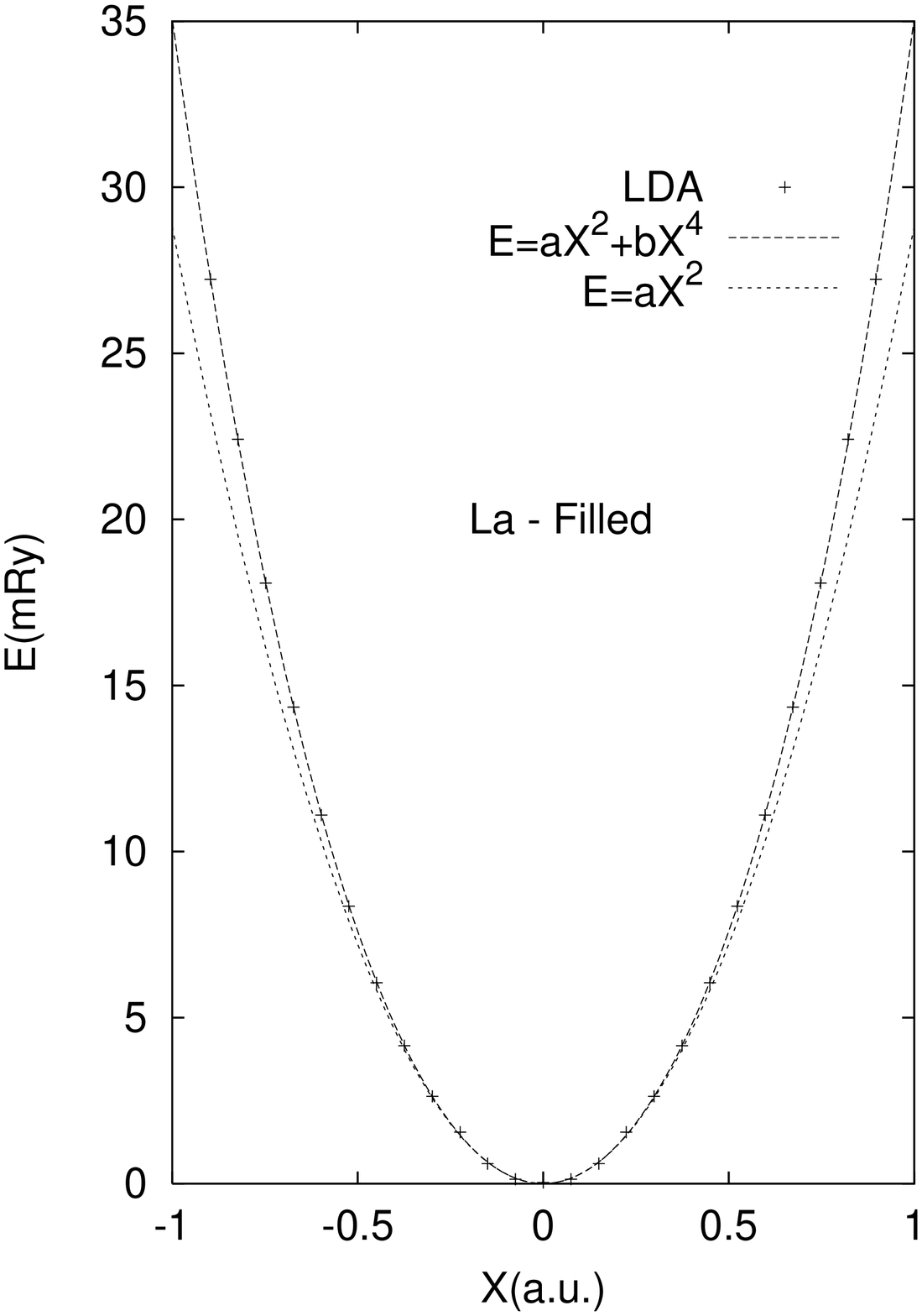,width=0.48\linewidth,height=.48\linewidth}}
\vspace{0.1in} \setlength{\columnwidth}{.96\linewidth} \nopagebreak
\caption{Calculated LDA energy vs. {\it RE} displacement along
a trigonal axis of {\it RE}Fe$_{4}$Sb$_{12}$. The least square fit
parameters are $a$=28.82 (24.64) mRy/bohr$^{2}$ and $b$=6.27 (5.50) mRy/bohr$%
^{4}$ for La (Ce).} \label{ener}
\end{figure}

Fig. \ref{ener} shows the LDA total energy as a
function of {\it RE} displacements in (La,Ce)Fe$_{4}$Sb$_{12}$: This
was calculated for a perfect crystal corresponding to the
skutterudite (bcc) lattice with displacements of the {\it RE}'s along a
trigonal axis. Not only is a double well excluded, but
the anharmonicity is of an ordinarily weak nature. In fact, addition
of a small
quartic anharmonic term is sufficient to fit the results to well within
0.1mRy over the entire ($\pm $0.5 \AA ) range. The
harmonic frequencies are 68 and 74 cm$^{-1}$ for Ce and La in the
corresponding materials and the anharmonic shift is less than a few cm$^{-1}$
at $T$ up to 1000K. (We refer to these frequencies as
``bare'' frequencies; we emphasize that they are neither crystalline
normal mode frequencies nor the ``single-ion'' frequency of
a {\it RE} with all other ions fixed {\it including} the
other {\it RE}'s). The crystal symmetry imposes isotropy on the harmonic
term but not on higher order terms. We computed the quartic parameter in the
La filled material with La displacement along a cubic axis to be 10.7mRy/bohr%
$^{4}$ as compared with 6.27mRy/bohr$^{4}$ for a trigonal axis.
These are the extrema of the quartic tensor.

\begin{figure}[tbp]
\centerline{\epsfig{file=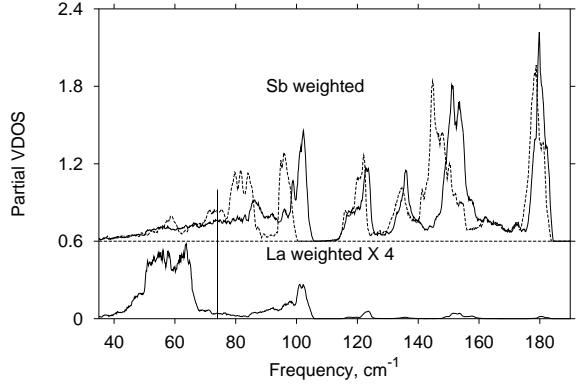,angle=-90,width=0.9\linewidth}}
\vspace{0.1in} \setlength{\columnwidth}{0.95\linewidth} \nopagebreak
\caption{La (bottom) and Sb (top) weighted
vibrational density of states (WVDOS) for LaFe$_4$Sb$_{12}$. The dashed line
shows Sb WVDOS for CoSb$_3$ which is similar to that for FeSb$_3$ within the
same dynamical model. The vertical line denotes the
``bare'' La frequency. The high frequency region where the Fe
atoms dominate is not shown.}
\label{dos} \end{figure}

Now we present the full dynamical model obtained from our
LDA results, including the above. We
added central force constants between the {\it RE} ion and its closest Fe
and Sb atoms to the valence force field model\cite{FS} for CoSb$_{3}$.
We found this simplest approximation to be quite sensible by
comparison with LDA forces for various sets of La, Sb, and Fe displacements.
Thus we directly get values of $-0.8$ and 1.64 (10$^{4}$
dyn/cm) for La-Fe and La-Sb force constants, respectively. The
{\it RE} bare frequency is then reproduced by construction. This
intermediate (${\cal I}
$)-model, based on a CoSb$_{3}$ model of Ref.\cite{FS} and
augmented by the above parameters to account for La filling, is useful in
analyzing the direct lattice dynamics effect of the RE. But, there is
also a considerable {\it indirec}t effect, by changes in other
force constants upon La filling and our final (${\cal F}
$)-model includes these, specifically a 30\% reduction
of the two central intra-square force constants and a 10\% reduction of
the M-Sb central force constants.
Below $\approx 140$ cm$^{-1}$ both models give similar
results. A test of the ${\cal F}$-model is to compare
frequencies of A$_{1g}$ phonons with those  calculated by the LDA for a Ce
comound\cite{NS} (137 and 157 cm$^{-1}$
): the ${\cal F}$-model gives 141 and 162 cm$^{-1},$ and the ${\cal I}
$-model 154
and 179 cm$^{-1}$.

\begin{figure}[t]
\centerline{\epsfig{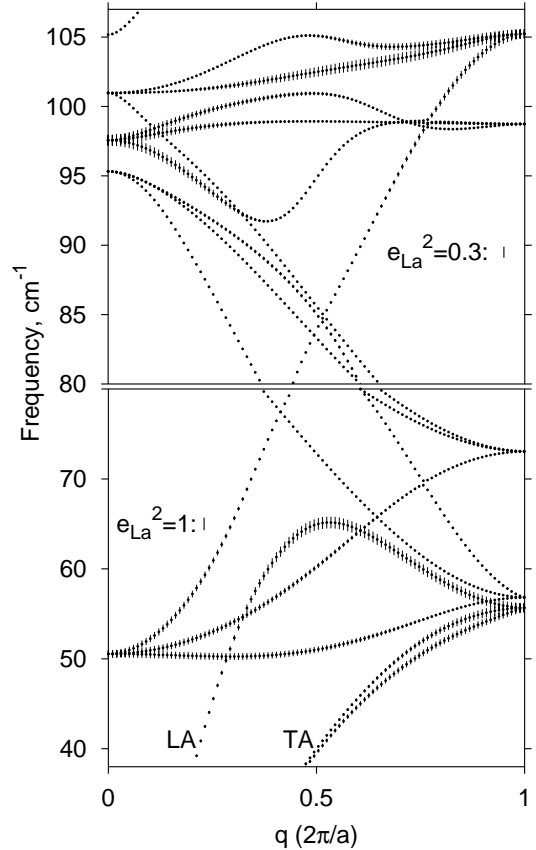}}
\vspace{0.1in} \setlength{\columnwidth}{0.95\linewidth} \nopagebreak
\caption{Dispersions along $\Gamma$-H. The widths
are the relative La character. Note the break in scale at 80 cm$^{-1}$}
\label{disp}
\end{figure}

The phonon dispersions of the unfilled material have
rather flat Sb derived optic branches quite close to the bare
frequency of 74 cm$^{-1}$ of the {\it RE} ions as seen in figures 4
through 7 of reference \cite{FS}. For example, one
observed IR
(F$_{u}$ symmetry) frequency in the unfilled structure is 78
cm$^{-1}$. F$_{u}$ symmetry allows for
{\it RE} displacements. We find strong {\it RE}-Sb coupling that
drives
predominantly Sb modes up in frequency and predominantly {\it RE}
modes down. Fig. \ref{dos} shows that the Sb
weighted VDOS is greatly affected by La-filling in the 80-100 cm$^{-1}$
region. With higher frequencies, or symmetries incompatible with La motion,
Sb-like peaks are also shifted up, as
the La-Sb force constant is positive, but the shifts are smaller. 
This is seen in the atom-type ($J$)
weighted vibrational density of states (Fig. \ref{dos}),
$G_{J}(\omega )=\sum_{ij}|{\bf e}
_{i}(j)|^{2}\delta (\omega -\omega _{i}),$ where the sum on $i$ is over
wave vectors and phonon branches and $j$ is over all type $J$ atoms in the
unit cell.{\sc \ }Figs. \ref{disp}a and b show the character of modes in the
filled material in the spectral region of interest. The dash lengths are
proportional to $|{\bf e}_{La}|^{2}$ where ${\bf e}_{La}$ is the
La-projection of the (normalized) polarization vector; {\it e.g.}, $
e_{La}^{2}=1$ would mean the mode involves only the La sublattice - this
would then have to be at the ``bare'' frequency of 74 cm$^{-1}$.
Strong coupling of Sb and {\it RE} motion is evident for all wavelengths.
A resonant interaction between a longitudinal
acoustic branch and a low lying ({\it RE} dominated) optic branch is found.
Fig. ~\ref{disp} shows the region of the acoustic spectrum where
resonant scattering by {\it RE} ions in disordered material can occur.
\cite{sethna} 

\begin{figure}[tbp]
\centerline{\epsfig{file=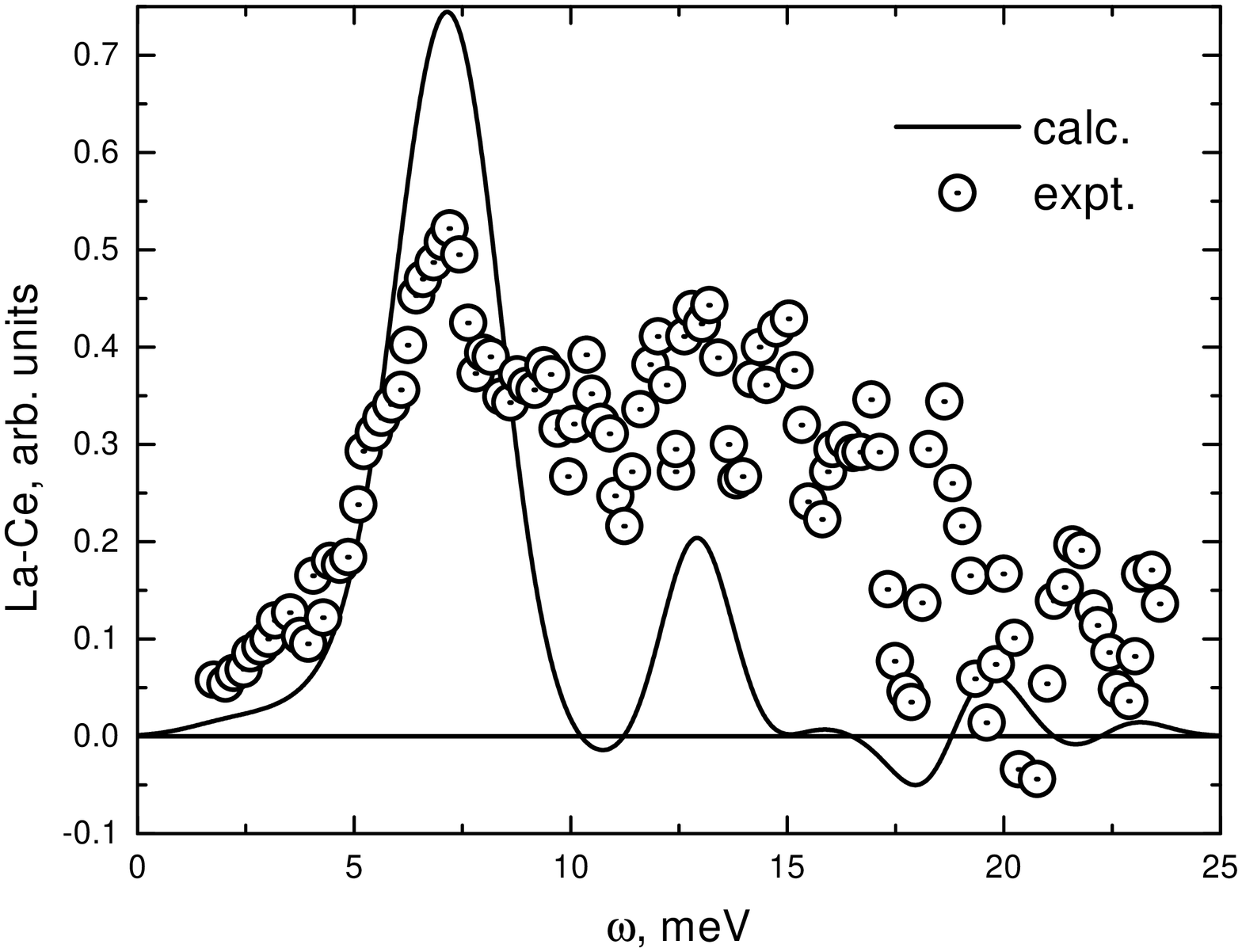,width=0.9\linewidth}}
\vspace{0.1in} \setlength{\columnwidth}{0.95\linewidth} \nopagebreak
\caption{INS spectral difference between La and Ce filled compounds.
Differences in the high frequency spectral region where the Fe and Co
atoms contribute heavily are not shown. Theoretical results are based on
model ${\cal F}$ (see text) and experimental results are from Ref. 
\protect\cite{kepp}}
\label{INS}
\end{figure}

Armed with this microscopic model, we can
discuss scenarios for LTC reduction and
related experiments, particularly inelastic
neutron scattering (INS) and specific heat (for T $\le$ 45 K)
for RECoFe$_{3}$Sb$_{12}$\cite{kepp}. A rather sharp peak
at 50 cm$^{-1}$ and a quite broad peak centered at about 100 cm$^{-1}$ were
obtained in a difference
spectrum, i.e., subtracting spectra
for La and Ce filled samples
and noting
that the cross section is much larger for La than for Ce,
to find {\it RE} vibrational contributions.\cite{kepp}
This led to a natural assertion of two distinct {\it RE} vibrations.

Using LDA calculations and our model, we can check
the main physical assumptions made in the earlier analysis.\cite{kepp}
First, it was assumed that there
is no difference in interatomic forces between La- and Ce-filled materials.
By LDA calculation we find that the self force constants of the La and
Ce do differ by 15\%, but, calculating the neutron
spectra from our model, we find that the La WVDOS is indeed
rather close to the difference in the INS between the  La- and
Ce-filled materials:
two major peaks, centered at 50 cm$^{-1}$ and 100 cm$^{-1}$, in the La weighted
VDOS (Fig. ~\ref{dos}),
are clearly seen
in the calculated differential INS spectrum (Fig. \ref{INS}) as well.
These correspond well to the two peaks found by
Keppens et al.\cite{kepp} (As usual, the experimental peaks are broader
due to instrumental and intrinsic materials effects like disorder and
anharmonicity.)
Most importantly, the calculated two peak structure was obtained
without an additional localized La mode.
Instead, the La spectral weight was transferred by hybridization
from the main La peak
(downshifted from the bare frequency of 78 cm$^{-1}$ to 50 cm$^{-1}),$ to Sb
modes at 100 cm$^{-1}.$

\begin{figure}[tbp]
\centerline{\epsfig{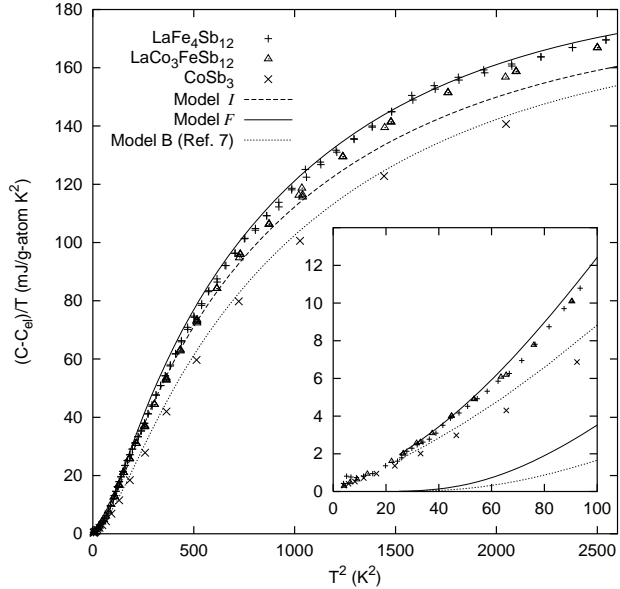}}
\vspace{0.1in} \setlength{\columnwidth}{0.95\linewidth} \nopagebreak
\caption{Comparison of specific heat measurements with the harmonic
force models for $T<50$K. The models, 1 and 2, for filled materials
are ${\cal I}$ and ${\cal F}$. The inset shows the large
effect of La modes in the T$\leq $10K region: The lower two curves were
calculated from the 50 to 70 cm$^{-1}$ spectral region exclusively.}
\label{cp1}
\end{figure}

Ref. \cite{kepp} presents more indications of two RE
frequencies (and possibly a TLS),
namely specific heat differences between CoSb$_{3}$
and RECoFe$_{3}$Sb$_{12}$ and small unusual temperature dependent features of
elastic constants. Clearly, a reconciliation of our
calculations that do not produce two separate La phonons, and the specific
heat experiments is needed. (We cannot address the temperature dependence
of the elastic constants in our calculations; 
this alone, however, cannot be considered as compelling  
evidence for TLS's.) Here we present new,
improved experimental
specific heat data that supersede the earlier data of Ref. \cite
{kepp}. The two sets agree below $T$=20K, but considerably disagree at higher
$T$. In terms of the previous Einstein modeling of the data for $T\alt50$ K, we
find that two frequencies are still needed to give the specific heat difference
between the unfilled and filled materials, but the strength
of the higher frequency component is much less.
The experiments were done on several compounds; CoSb$_{3}$, LaCoFe$_{3}$%
Sb$_{12}$ and LaFe$_{4}$Sb$_{12}$. The synthesis procedure was
previously described\cite{sales}. Heat capacity data from 2-300 K were
taken with a commercial Quantum Design system. Results from this
system are in good agreement with published values for
standards like sapphire and copper. Some of the new results are shown in
Fig. ~\ref{cp1}\cite{sh} along with our calculations.
The new experimental heat capacity for the filled
material is much closer to that of the unfilled material than the
earlier experiment.

The agreement with the present theoretical results for C$_{V}$
is remarkable, especially considering that the model does not have the two
additional localized high-frequency modes.
Resolution of this seeming paradox lies in an implicit
assumption, which lacking force constant information had to be made
in Ref.\cite{kepp}, i.e. that interatomic forces are
transferable between the filled and unfilled materials and that specific
heat differences between the materials are solely from {\it RE}
vibrations. However,
our LDA calculations show an important contribution to the Sb
dynamics from La-Sb forces; besides, the intra-square Sb
interactions themselves are decreased by
approximately 30\% upon filling. While new rather localized La modes do
appear upon filling in our analysis (at $\approx 50$ cm$^{-1}$), and they are 
reflected in the increased specific heat at the low temperatures,
there are two more 
changes which greatly affect the specific heat at low temperatures:
first, as discussed,
certain Sb modes are shifted to higher frequency, which decreases the
specific heat; second, the softening of the Sb-Sb
forces increases the specific heat.
At still higher temperature (50--300 K)
the lattice specific heat approaches the harmonic Dulong-Petit value:
There is good agreement with the models, but those measurements are not
shown here.

So the origin of the LTC suppression is plainly more subtle than
prior speculations about strongly anharmonic {\it RE} motions.
Aspects of our results suggest scenarios for
the role of the {\it RE} ions: (1) There is
significant harmonic interaction between La and Sb in comparison to the
inter-square Sb force constants and most likely this holds for the
anharmonic interactions as well\cite{cnote}.
Thus the rare earth vibrations, which are
not heat carriers due to their flat dispersion, do interact
strongly with heat carrying phonons. Slack
and Galginaitis\cite{slga}
pointed out that the effect of Raman-like scattering of phonons by
magnetic impurity levels in CdTe materials could reduce LTC
and the reduction would be greater with increasing temperature.
Perhaps a similar effect takes place here via the cubic anharmonic
interaction between heat carrying phonons and the 50 cm$^{-1}$ predominantly
La vibrations even for a perfect crystal. (2) Typically there is a
significant lack of complete rare earth
filling in samples. Harmonic resonant scattering of phonons by ''impurity''
La (or La vacancy)
vibrations could give short relaxation times for frequencies
near the La modes. These are
in a region of high diffusivity, $Gv^{2}$
in the models. Also,
the reduction in Sb-Sb interactions on filling
implies force constant disorder in the Sb sublattice of partly
filled samples (lattice specific heats of LaFe$_{4}$Sb$%
_{12}$ and LaCoFe$_{3}$Sb$_{12}$ are extremely similar in the T$\le$50K
region so the
reduction in Sb-Sb interactions seems due to {\it RE}
filling, not substitution of Co by Fe).

To summarize, we find that the {\it RE}'s are in a harmonic well
up to large displacements, implying that the simplest rattling ion models
are inapplicable to these filled skutterudites; we
explain the two La peak spectrum by harmonic
lattice dynamics as follows: Without the ${\it dynamical}$ La-Sb interaction
the pure La modes would be concentrated around 70 cm$^{-1}$,
with a substantial number of Sb modes at slightly higher
frequencies. Strong La-Sb hybridization pushes these groups apart,
to $\approx $ 50 and $\approx $ 100 cm$^{-1}$.
We suggest, on the basis of our
calculations, possible mechanisms for {\it RE} induced LTC reduction.

ONR supported work at NRL. We thank V. Keppens, G. Slack, 
D. Morelli, J.-P. Fleurial, and T. Caillat for helpful conversations.
Computations were performed at the DoD ASC center.

\multe


\begin{references}
\bibitem{sh}  Electronic contributions (substantial only for filled
materials) have been extracted from the low-T data and subtracted out
at all temperatures assuming linear temperature dependence.
The linear coefficients are 10 and 3.4 mJ/g-at. K$^{2}$
for the Fe and Fe/Co materials, respectively. We neglect the C$_{P}$-C$_{V}$
correction as we estimated it to be only of the order 0.15 J/gm-atom
K at 300K, less than the uncertainty in the electronic specific heat.
\bibitem{cnote}
We did some LDA calculations to assess the
anharmonic La-Sb coupling; the ratio of the corresponding cubic
parameter to the harmonic is 0.7 of
that for a Lennard Jones potential at the potential minimum.
\end{references}
\end{document}